\documentstyle[12pt,a4]{article}
\begin{document}
\title{DOES THE G\"{U}RSEY--TZE SOLUTION REPRESENT A MONOPOLE CONDENSATE?}
\author{Serdar Nergiz and Cihan Sa\c{c}l\i o\~{g}lu\thanks{Permanent addresss:
Physics Department, Bo\~{g}azi\c{c}i University,
80815 Bebek--\.{I}stanbul, Turkey} }
\date{Physics Department, Bo\~{g}azi\c{c}i University \\
80815 Bebek--\.{I}stanbul, Turkey \\  and \\ Physics Department, TUBITAK
\\  Marmara Research Center \\
Research Institute for Basic Sciences \\ 41470 Gebze, Turkey}
\maketitle
\vspace*{1cm}
\begin{abstract}
We recast the quaternionic G\"{u}rsey-Tze solution, which is a fourfold
quasi-periodic self-dual Yang-Mills field with a unit instanton number per
Euclidean spacetime cell, into an ordinary coordinate formulation. After
performing the sum in the Euclidean time direction, we use an observation
by Rossi which suggests that the solution represents an arrangement with
a BPS monopole per space lattice cell. This may provide a concrete realization
of a monopole condensate in pure Yang-Mills theory.
\end{abstract}
\vspace*{5 cm}
\pagebreak
\baselineskip=30pt

Recently, Seiberg and Witten \cite{art1,art2,art3} were able to verify
explicitly that the condensation of magnetic monopoles results in the
confinement of Yang-Mills electric charge. Their analysis, however, involves
the use of additional fields provided by $N=2$ supersymmetry and thus leaves
the question of how the monopoles are to arise in pure Yang-Mills theory
unanswered.

In this note, we take as our starting point an old idea
of Julia and Zee \cite{art4} and Rossi
\cite{art5} , who pointed out that a static Euclidean $A^{a}_{0}$ may be
identified with the Higgs field $\varphi^{a}$ in the BPS \cite{art6} limit.
The self-dual Yang-Mills equations are then formally identical with the
Bogomolny equations expressing the proportionality between the mass and the
magnetic charge of the monopole. Rossi showed in particular that a sequence
of equal size Jackiw-Nohl-Rebbi instantons \cite{art7} arranged periodically
along the Euclidean time axis is gauge-equivalent to a BPS monopole with the
above identification. The mass of the monopole turns out to be the action per
unit time; it is thus inversely proportional to the separation
between the instantons.

Now let us choose a second line parallel to the Euclidean time axis and
place instantons on it at the same time locations as on the first. Since
instanton numbers are simply additive, this will represent a solution with
twice the action density or ``mass'' as in the BPS case, or, in other words, a
separated two-monopole solution, albeit in a gauge where an artificial
periodic time dependence is present. The time dependence can in fact easily
seen to become negligible at large distances from the centers of the solutions.

The strategy for obtaining a solution with BPS monopoles arranged on a space
lattice (which may be regarded as a realization of the monopole condensate)
then would seem to be to repeat Rossi's argument for every lattice point in
space, or, in other words, to find a fourfold periodic instanton configuration.
Interestingly, the Copenhagen vacuum \cite{art8} corresponds to a doubly
periodic array of Nielsen-Olesen vortices on, say, the $xy$-plane. If one
could extend Rossi's argument one more step by showing the equivalence
of BPS monopoles arranged periodically along the $z$-axis to a Nielsen-Olesen
vortex, the Copenhagen vacuum could then be viewed as an alternative
description of a monopole condensate or a fourfold periodic arrangement of
instantons. It is, of course, to be kept in mind that such classical solutions
can only model the vacuum over a limited domain, with quantum fluctuations
restoring Lorentz and rotational invariance in an average sense \cite{art8}.
Furthermore, the most general ADHM solution \cite{art9} not being available
in explicit form, one has to be content with the next most general, namely
the Jackiw-Nohl-Rebbi version, of such an infinite-instanton configuration.

If one adopted the above line of reasoning, it would seem that the connection
\begin{equation}\label{1}
A_{\mu} = i \overline{\sigma}_{\mu \nu} \partial_{\nu} \ln{\rho}
\end{equation}
with
\begin{equation}\label{2}
\rho = \sum_{n_{0}} \sum_{n_{1}} \sum_{n_{2}} \sum_{n_{3}} \frac{1}{(x-q)^{2}}
\end{equation}
where
\begin{equation}\label{3}
q_{\mu} = n_{0} q_{\mu}^{(0)}+n_{1} q_{\mu}^{(1)}+n_{2} q_{\mu}^{(2)}+
n_{3} q_{\mu}^{(3)}
\end{equation}
would immediately provide the desired solution. In the above, the
$\overline{\sigma}_{\mu \nu}$
are the Pauli matrices corresponding to the `t~Hooft symbols \cite{art10},
the
$ q^{(a)}$ ($a= 0,1,2,3$)
are the lattice vectors and the
$n_{a}$
are integers ranging from minus to plus infinity. However, (\ref{2}) cannot
be accepted as it stands because of the divergent quadruple sum:
Comparing the sum with an integral over q, we find the behavior
$\int d|q| \, |q|^{3} / |q|^{2}$ .
Thus subtraction terms are called for.

This is, of course, familiar from the definition of Weierstrassian elliptic
functions: The subtraction terms in
\begin{equation}\label{4}
\wp (z) \equiv \frac{1}{z^{2}} +
\sum_{\omega \neq 0} \{ \frac{1}{(z-\omega)^{2}}
-\frac{1}{\omega^{2}} \} ~~,~~ (\omega = n_{1} \omega_{1} + n_{2} \omega_{2} )
\end{equation}
and
\begin{equation}\label{5}
\zeta (z) \equiv - \int \wp (z) dz = \frac{1}{z} + \sum_{\omega \neq 0}
\{ \frac{1}{(z-\omega)} + \frac{1}{\omega} + \frac{z}{\omega^{2}} \}
\end{equation}
produce the required convergence. Note the inviolability of the parentheses:
the
$ 1 / \omega$
term would sum to zero by itself if the terms in brackets could be
considered separately. A safer way of writing (\ref{5}) is
\begin{equation}\label{6}
\zeta (z) = \frac{1}{z} + \sum_{\omega \neq 0}
\frac{z^{2}}{\omega^{2}(z-\omega)}
\end{equation}
where now the convergence is manifest but the (quasi) periodicity is less
obvious than in the expression (\ref{5}). It is of course only
$\wp (z)$
that is doubly periodic while
$\zeta (z)$
obeys the quasi-periodic
transformation law
\begin{equation}\label{7}
\zeta (z + \omega_{1,2}) = \zeta (z) + \eta_{1,2}
\end{equation}
under shifts by the lattice vectors
$\omega_{1}$
and
$\omega_{2}$.
The
constant complex numbers
$\eta_{1}$
and
$\eta_{2}$
are related to the periods
$\omega_{1}$
and
$\omega_{2}$
through Legendre's relation
\begin{equation}\label{8}
\eta_{1} \omega_{2} - \eta_{2} \omega_{1} = 2 \pi i = \oint_{\partial \, \rm
cell} \zeta (z) dz ~~.
\end{equation}

The question is then how one should modify (\ref{2}) in analogy to (\ref{4})
and (\ref{5}) in order to attain convergence. The general strategy is clear:
Three subtraction terms are needed, the first being
$- 1/q^{2} $.
On dimensional grounds, the next two should have the form
$x/q^{3} $
and
$x^{2}/q^{4} $.
What is not clear is exactly how these last two terms are to be chosen.

The answer has fortunately been provided by R.~Fueter \cite{art11} and
already exploited by G\"{u}rsey and Tze \cite{art12}, whose results and
techniques we now summarize to keep the presentation self-contained.
Introduce the unit quaternions
$ \bf e_{\mit 0} \mit = I $
 and
$ \bf e_{\mit i} $~$(i=1,2,3)$
with
$ \bf e_{\mit i} \bf e_{\mit j} \mit = - \delta_{ij} +
\epsilon_{ijk} \bf e_{\mit k} $
and associate a quaternion
$ \bf v \mit= v_{\mu} \bf e_{\mu} \mit $
with the Euclidean 4-vector
$v_{\mu}$.
The conjugate
$ \overline{\bf v \mit} $
of
$ \bf v \mit $
is defined as
$ \overline{\bf v \mit} =  v_{\mu} \overline{\bf e}_{\mit \mu} $,
where
$ \overline{\bf e}_{\mit \mu}  = ( \bf e_{\mit 0} \mit ,- \vec{\bf v \mit} ) $.
One also defines the Dirac-like operators
$ D = \bf e_{\mu} \mit \partial_{\mu} $
and
$ \overline{D} = \overline{\bf e}_{\mit \mu} \partial_{\mu} $
which obey
$ D \overline{D} = \overline{D} D = \Box $.
In this notation, Fueter's Z-function which looks superficially similar
to (\ref{5}) reads
\begin{equation}\label{9}
Z(\bf x \mit) = \frac{1}{\bf x \mit} + \sum_{\bf q \mit \neq 0}
\{ \; \frac{1}{\bf x-q \mit} +
\frac{1}{\bf q \mit} +
\frac{1}{\bf q \mit} \bf x \mit \frac{1}{\bf q \mit} +
\frac{1}{\bf q \mit} \bf x \mit \frac{1}{\bf q \mit} \bf x \mit
\frac{1}{\bf q \mit} +
\frac{1}{\bf q \mit} \bf x \mit \frac{1}{\bf q \mit} \bf x \mit
 \frac{1}{\bf q \mit} \bf x \mit \frac{1}{\bf q \mit} \; \}~~,
\end{equation}
while the manifestly convergent form similar to (\ref{6})becomes
\begin{equation}\label{10}
Z(\bf x \mit) = \frac{1}{\bf x \mit} + \sum_{\bf q \mit \neq 0}
(~ \frac{1}{\bf q} \bf x \mit ~)^{4} \frac{1}{\bf x-q \mit} ~~.
\end{equation}
The Z-function is however not the true analog of
$ \zeta (z) $
since it does not obey a quasiperiodicity relation analogous to (\ref{7});
it is rather the function
\begin{equation}\label{11}
\zeta^{F} (\bf x \mit) \equiv \Box Z(\bf x \mit)
\end{equation}
which satisfies
\begin{equation}\label{12}
\zeta^{F} (\bf x \mit + \bf q^{(\mit a)}) \mit = \zeta^{F} (\bf x \mit) +
\mbox{\boldmath $\eta$ \unboldmath }^{(\mit a)} ~~,
\end{equation}
where the
\boldmath $ \eta^{(a)} $ \unboldmath
are constant quaternions. The analog of Legendre's relation is
\begin{equation}\label{13}
\sum_{a=0}^{3} \mbox{\boldmath $\eta$ \unboldmath }^{(\mit a)}
\bf Q^{(\mit a)} \mit = 8 \pi^{2}~~,
\end{equation}
where
\begin{equation}\label{14}
\bf Q^{(\mit a)} = e_{\mu} \mit \epsilon_{\mu \nu \alpha \beta} q_{\nu}^{(b)}
q_{\alpha}^{(c)} q_{\beta}^{(d)} ~~(a, b, c, d ~ cyclic).
\end{equation}
Note that (\ref{13}) is different from equation~(6.37) in reference~
\cite{art12}; that (6.37) requires correction is obvious on dimensional
grounds. G\"{u}rsey and Tze use the fact that
$ \Box DZ(x) = 0 $
and thus propose
\begin{equation}\label{15}
\rho = DZ
\end{equation}
in place of (\ref{2}).

One of our main results is that the G\"{u}rsey-Tze solution can be
written in a quaternion-free form where its spacetime structure becomes
more transparent. After somewhat lengthy manipulations using quaternionic
algebra and differentiation, one finds
\begin{equation}\label{16}
\rho (x) = \frac{2}{x^{2}} + 2 \sum_{q \neq 0} \{
\frac{1}{(x-q)^{2}} - \frac{1}{q^{2}} - \frac{2 q \cdot x}{q^{4}} -
\frac{1}{q^{6}} (4(q \cdot x)^{2} - q^{2} x^{2}) \} ~.
\end{equation}
We will not be using the quaternions
$ \bf x \mit $
any longer; in (\ref{16}) and the following, we revert to standart
4-vector notation. Note that the first two terms are just (\ref{2});
the third and fourth terms are obviously harmonic and the
$ 1/q^{6} $
term is easily shown to be harmonic. The terms added to (\ref{2}) are such
that for
$ x^{2} \ll q^{2} $,
(\ref{16}) becomes
\begin{equation}\label{17}
\rho (x) \cong \frac{2}{x^{2}} + 2 \sum_{q \neq 0}
( \frac{x^{4}}{q^{6}} - \frac{12}{q^{8}} x^{2} (q \cdot x)^{2} +
\frac{16}{q^{10}} (q \cdot x)^{4} ) +
O( \frac{x^{6}}{q^{8}} ) ~...~,
\end{equation}
where even powers of
$x$
between
$x^{-2}$
and
$x^{4}$
are seen to be absent.

The reader may wonder how the quaternionic analogue of
$ \wp (z) $
might be defined. While this is not explicitly defined in \cite{art12}  ,
a natural definition appears to be
$ \wp^{F} (x) = D\zeta = D\bar{D} \rho = - 2 \pi^{2} \sum_{\rm all \: \mit q}
\delta^{4} (x-q) $,
which is a distribution rather than a function! Note that this
``$\, \wp^{F} (x) \,$''
retains no information from
$ \rho (x) $
except for the locations of the instantons.

It is again tempting but wrong to sum the terms in the curly brackets in
(\ref{16}) separately. For example, the
$ 2 q \cdot x / q^{4} $
term, which, if added naively, would give zero just like the
$1/ \omega $
term in (\ref{5}), is only meaningful in conjunction with the other
terms. Similarly, the
$1/q^{6} $
term which superficially appears to vanish for a cubic lattice should be
kept along with the others regardless of the lattice type. The
manifestly convergent form of (\ref{16}) is
\begin{equation}\label{18}
\rho (x) = \frac{2}{x^{2}} + \sum_{q \neq 0}
\frac{16 (q \cdot x)^{3} - 8 q^{2} x^{2} (q \cdot x) +
2 q^{2} x^{4} - 8 x^{2} (q \cdot x)^{2}}{q^{6} (x-q)^{2}}
\end{equation}
where the sum on each term may now be separately performed. It is safe to
conclude from (\ref{18}) that
$ \rho (-x) = \rho (x) $
by changing
$q$
to
$-q$
in the sum wherever necessary. Using this evenness property together
with
$ \zeta^{F} = \overline{D} \rho $
and (\ref{12}), we can find the behavior of
$ \rho $
under lattice displacements. We first strip (\ref{12}) of the
quaternion units
$ \bf e_{\mu} \mit $
to get
\begin{equation}\label{19}
\partial_{\mu} \rho (x + q^{(a)}) = \partial_{\mu} \rho (x) +
\overline{\eta}_{\mu}^{(a)}~~,
\end{equation}
where
$ \overline{\eta}_{\mu}^{(a)} = ( \eta_{0}^{(a)} , - \vec{\eta}^{\, (a)} ) $.
An integration gives
\begin{equation}\label{20}
\rho (x + q^{(a)}) = \rho (x) + \overline{\eta}^{(a)} \cdot x + c~~,
\end{equation}
with
$c$
an integration constant. Putting
$ x = - q^{(a)}/2 $
and using the evenness of
$ \rho (x) $,
we find
\begin{equation}\label{21}
c = \frac{1}{2} \overline{\eta}^{(a)} q^{(a)}
\end{equation}
where \underline{no} sum over the index
$a$
is implied. The transformation of the connection (\ref{1}) under a
lattice shift is thus dictated by (\ref{20}) and (\ref{21}). It is
not difficult to see that even gauge invariant quantities such as the
Lagrangian density change under this transformation; however, each
spacetime cell contributes unit topological charge to the action
\cite{art12}. Adopting Rossi's interpretation of the BPS monopole as
action per unit time, the same topological charge per spacetime cell can
also be viewed as a magnetic monopole density in space.

It is useful to perform the sum over the Euclidean time axis in the
equation (\ref{16}) and to compare it with Rossi's result. Note that
\underline{one} single sum, being convergent, may be performed
\underline{separately} on the individual terms in (\ref{16}) although
subsequent such sums are not allowed. It is much harder to sum the
manifestly convergent expression (\ref{18}). The basic technique is the
Sommerfeld-Watson transform. Let us first rewrite Rossi's result in the
form
\begin{equation}\label{22}
\rho = \sum_{- \infty}^{\infty} \frac{1}{( r^{2} + (t - n \tau )^{2} )}
= \frac{i \pi}{2 \tau r} \{
\frac{1}{\tan \frac{ \pi}{\tau} (t + i r ) } -
\frac{1}{\tan \frac{ \pi}{\tau} (t - i r ) } \}~.
\end{equation}

The gauge transformation which brings this into the static BPS form
is given by
\begin{equation}\label{23}
U(\theta) = \exp (- i \vec{T} \cdot \hat{n} \theta )
\end{equation}
where
\begin{equation}\label{24}
\theta = \arctan \{ \frac{\sin \frac{2 \pi t}{\tau} ~
\sinh \frac{2 \pi r}{\tau}}{\cos \frac{2 \pi t}{\tau} ~
\cosh \frac{2 \pi r}{\tau} -1 } \}~.
\end{equation}
The monopole mass is
\begin{equation}\label{25}
\frac{dS}{dt} = \frac{8 \pi^{2}}{\tau g^{2}}~~,
\end{equation}
where
$g$
is of course the coupling constant.

While spotting the BPS monopole in the pre-gauge transformed
expression (\ref{22}) is far from trivial, the infinite mass limit
obtained by
$ \tau \rightarrow 0 $
provides a quick insight into the static monopole nature of the
solution, albeit in singular form. One finds
\begin{equation}\label{26}
\lim_{\tau \rightarrow 0} \rho = \frac{\pi}{\tau r}~,
\end{equation}
which gives
\begin{equation}\label{27.a}
A_{i}^{a} = \epsilon_{iab} \frac{x_{b}}{r^{2}}
\end{equation}
and
\begin{equation}\label{27.b}
A_{0}^{a} = \frac{x_{a}}{r^{2}} ~,
\end{equation}
where the fields are seen to have the required orientation in
ordinary and
$ SU(2) $
space. We will apply the same limit on our result later.
In doing a similar sum over (\ref{16}), we have to observe the
restriction that not all the
$ ( n_{0}, \:n_{1}, \:n_{2}, \:n_{3} ) $
in (\ref{3}) are allowed to vanish simultaneously. We can take this
into account by splitting the sum into two parts:
\begin{eqnarray}\label{28}
\frac{1}{x^{2}} + (\sum_{n_{3}} \sum_{n_{2}} \sum_{n_{1}}
\sum_{n_{0}})^{\prime}
 & = &
\frac{1}{x^{2}} +
(\sum_{\begin{array}{rcl}
\scriptstyle n_{\scriptscriptstyle 0} \! \! & \! \! \scriptstyle \! \! =
\! \! & \! \! \scriptstyle \! \! - \scriptstyle \infty \\
\scriptstyle \vec{\scriptstyle n}     \! \! & \! \! \scriptstyle \! \! =
  \! \! & \! \! \scriptstyle \! \! 0
           \end{array} }^{-1} +
\sum_{\begin{array}{rcl}
\scriptstyle n_{\scriptscriptstyle 0} \! \! & \! \! \scriptstyle \! \!
 =  \! \! & \! \! \scriptstyle \! \! 1 \\
\scriptstyle \vec{\scriptstyle n}     \! \! & \! \! \scriptstyle \! \!
=  \! \! & \! \! \scriptstyle \! \! 0
        \end{array}}^{\infty} ) +
( \sum_{} \sum_{ |\vec{n}| \neq 0 } \sum_{} ) \sum_{\rm all \: \mit n_{0}}
\nonumber \\
\nonumber \\
& = &
\frac{i \pi}{\tau r}
\{
\frac{1}{\tan \frac{\pi}{\tau} (t + i r)} -
\frac{1}{\tan \frac{\pi}{\tau} (t - i r)}
\}
- \frac{2 \pi^{2}}{3 \tau^{2}} +
\frac{2 \pi^{4}}{45 \tau^{4}} (x^{2} - 4 t^{2})
\nonumber \\
&  & \mbox{}
+ \sum_{} \sum_{ |\vec{n}| \neq 0 } \sum_{} [
\frac{i \pi}{\tau |\vec{r}-\vec{q} \, | }
\{
\frac{1}{\tan \frac{\pi}{\tau} (t + i | \vec{r} - \vec{q} \, |)} -
\frac{1}{\tan \frac{\pi}{\tau} (t - i | \vec{r} - \vec{q} \, |)}
\}
\nonumber \\
&  & \mbox{}
- \frac{2 \pi}{\tau |\vec{q} \, |} \coth \frac{\pi |\vec{q} \, |}{\tau} +
\frac{\pi (x^{2} - 2 \vec{q} \cdot \vec{r} )}{\tau |\vec{q} \, |^{2}}
\{
\frac{1}{|\vec{q} \, |} \coth \frac{\pi |\vec{q} \, |}{\tau} +
\frac{\pi}{\tau} \frac{1}{\sinh^{2} \frac{\pi |\vec{q} \, |}{\tau}}
\}
\nonumber\\
&  & \mbox{}
- \frac{\pi (\vec{q} \cdot \vec{r})^{2}}{\tau |\vec{q} \, |^{3}}
\{
\frac{3}{|\vec{q} \, |^{2}} \coth \frac{\pi |\vec{q} \, |}{\tau} +
\frac{3\pi}{\tau|\vec{q} \, |} \frac{1}{\sinh^{2}
\frac{\pi|\vec{q} \, |}{\tau}}+
\frac{2\pi^{2}}{\tau^{2}}
\frac{\cosh \frac{\pi |\vec{q} \, |}{\tau}}{\sinh^{3} \frac{\pi|\vec{q} \,
|}{\tau}}
\}
\nonumber\\
&  & \mbox{}
- \frac{\pi t^{2}}{\tau |\vec{q} \, |^{2}}
\{
\frac{1}{|\vec{q} \, |} \coth \frac{\pi |\vec{q} \, |}{\tau} +
\frac{\pi}{\tau} \frac{1}{\sinh^{2} \frac{\pi |\vec{q} \, |}{\tau}} -
\frac{2\pi^{2} |\vec{q} \, |}{\tau^{2}}
\frac{\cosh \frac{\pi |\vec{q} \, |}{\tau}}{\sinh^{3} \frac{\pi |\vec{q}
\, |}{\tau}}
\} ]
\end{eqnarray}

Comparing (\ref{28}) with (\ref{22}), one notices the first two terms in the
triple sum, which have the appearance of magnetic monopoles centered at
the lattice sites
$ \{ \vec{q}\, \} =
\{ n_{1} \vec{q}^{\, (1)} + n_{2} \vec{q}^{\, (2)} + n_{3}
\vec{q}^{\, (3)} \} $.
The remaining terms are needed to make the sum convergent. Next,
consider the
$ \tau \rightarrow 0 $
limit of (\ref{28}), which gives
\begin{eqnarray}\label{29}
\lim_{\tau \rightarrow 0}
\{
\frac{1}{x^{2}} +
(\sum_{n_{3}} \sum_{n_{2}} \sum_{n_{1}} \sum_{n_{0}})^{\prime}
\}
& = &
\frac{2\pi}{\tau}
\{
\frac{1}{r} -
\frac{\pi}{3\tau} +
\frac{\pi^{3}}{45\tau^{3}} (x^{2}- 4 t^{2}) +
\sum_{} \sum_{ |\vec{n}| \neq 0 } \sum_{}
[\frac{1}{| \vec{r} - \vec{q} \, |}
\nonumber \\
&  & \mbox{}
- \frac{1}{|\vec{q} \, |} +
\frac{r^{2} - 2 \vec{q} \cdot \vec{r}}{2|\vec{q} \, |^{3}} -
\frac{3(\vec{q} \cdot \vec{r})^{2}}{2|\vec{q} \, |^{5}} ]
\}
\end{eqnarray}
The analogs of singular monopoles (as in (\ref{27.a} - \ref{27.b} )) at
the sites
$ \vec{q} $
again make their appearance. The third term is an artifact of the way the
sum was split in (\ref{28}); it can be made to disappear if single sums
over
$ n_{1} $,
$ n_{2} $,
and
$ n_{3} $
are separated out in the same way the
$ n_{0} $
sum was singled out in (\ref{28}); however, this will introduce time
dependence in other terms. Unlike in the Rossi case, the solution
(\ref{16}) treats all coordinates alike; hence it should not come as a
surprise that the time dependence cannot be removed. Were it removable,
one could also extend this to the other coordinates and obtain a constant
solution, contradicting the nontrivial coordinate dependence evident in
(\ref{16}).

The manageable and legitimate sums stop here; however, if we ignore the
convergence problem, assume a square lattice in
$ ( q^{(0)} , q^{(3)} ) $
and do a naive integration over the variable
$ q_{3} $
for the terms
$ | \vec{r} - \vec{q} \, |^{-1}  $
and
$ - | \vec{q} \, |^{-1}  $,
we obtain expressions like
$ \rho \sim \ln (| \vec{r}_{\perp} - \vec{q}_{\perp}  | /
|\vec{q}_{\perp}  |) + \cdots ~, $
which are singular versions of non-Abelian Nielsen-Olesen vortices
\cite{art13,art14} centered at the locations
$ | \vec{q}_{\perp} | = n_{1} \vec{q}^{\, (1)} + n_{2} \vec{q}^{\, (2)} $~.
This perhaps provides some support for the idea, mentioned in the beginning
of this paper, that a fourfold quasi-periodic solution and the Copenhagen
vacuum may be related.

So far, we have not made any specific choices for the lattice generated by
the
$ q^{(a)} $.
As long as no detailed dynamical calculations are attempted, there seems
to be no reason to prefer one lattice over another. However, the situation
may be different if, say, the vacuum energy density is calculated to one
loop: For example, when such a calculation is done for various trial
Copenhagen vacua \cite{art8} , the hexagonal
$ SU(3) $
root lattice is seen to be energetically favored. While we have not yet
carried out a similar investigation, it would be surprising if the root
lattice of
$ SO(8) $ ,
corresponding to the tightest packing of spheres in four dimensions, were
not found to play a distinguished role in our problem. In fact, the
preferred
$ SU(3) $
root lattice in \cite{art8} is a sublattice of the
$ SO(8) $
root lattice. Curiously,
$ SO(8) $
also makes an appearance in \cite{art3}, where it is mixed with
electric-magnetic duality.

{\bf Acknowledgements}

We thank our Theory Group colleagues Gebze and Bo\~{g}azi\c{c}i for useful
discussions and encouragement. C. S. gratefully acknowledges the hospitality
and support extended to him at the Marmara Research Center.
This work was partially
supported by the Turkish Scientific and
Technical Research Council, TBAG \c{C}G-1.

\end{document}